\documentclass[]{spie}  

\usepackage{amsmath,amsfonts,amssymb}
\usepackage{graphicx}
\usepackage[colorlinks=true, allcolors=blue]{hyperref}

\title{Characterizing, correcting, and repairing the effects of radiation damage in the COSI germanium cross-strip detectors}

\author[a,b]{Steven E. Boggs}
\author[a]{Sophia E. Haight}
\author[a]{Sean N. Pike}

\author[a]{Jarred Roberts}
\author[c]{Albert Y. Shih}
\author[d]{Joanna M. Szornel}
\author[b]{John A. Tomsick}
\author[b]{Andreas Zoglauer}

\affil[a]{Department of Astronomy and Astrophysics, University of California, San Diego, 9500 Gilman Drive, La Jolla, CA, 92093, USA}
            
\affil[b]{Space Sciences Laboratory, University of California, Berkeley, 7 Gauss Way, Berkeley, CA, 94720, USA}
            
\affil[c]{NASA Goddard Space Flight Center, Greenbelt, MD, 20771, USA}

\affil[d]{Lawrence Berkeley National Laboratory, 1 Cyclotron Road, Berkeley, CA, 94720, USA}

\authorinfo{Further author information: (Send correspondence to S.E.B.)\\S.E.B.: E-mail: seboggs@ucsd.edu}

\begin{document} 
\maketitle

\begin{abstract}
The Compton Spectrometer and Imager (COSI) is a gamma-ray survey telescope utilizing a compact Compton imager design, enabled by an array of 16 high-resolution germanium cross-strip detectors. After its launch into an equatorial Low Earth Orbit (LEO) in 2027, COSI will experience radiation damage primarily due to energetic protons, with the proton fluence dominated by the passage of COSI through the edge of the South Atlantic Anomaly (SAA) for a few minutes each orbit. We have developed a comprehensive program focused on the  modeling, characterization, data correction, and physical repair of radiation damage effects in the COSI detectors. We have performed energetic proton beam irradiations of a spare COSI detector at a proton synchrotron, with proton fluences consistent with multiple years of exposure to the COSI space radiation environment. These exposures allow us to characterize the relationship between proton fluence and induced charge trapping. We demonstrate our techniques to correct for trapping effects, as well as characterize the effectiveness of high-temperature annealing on correcting this damage, as characterized by the resulting spectral performance of the detector. We will present our efforts to characterize the effects of radiation damage in the COSI detectors, as well as our techniques for correcting these effects in the data analysis pipeline and ultimately repairing the detectors on orbit every few years through high-temperature annealing. 
\end{abstract}

\keywords{Germanium semiconductor detectors; Charge trapping; Gamma-ray spectroscopy; Radiation damage; Physics - Instrumentation and Detectors; Astrophysics - Instrumentation and Methods for Astrophysics}

\section{INTRODUCTION}
\label{sec:intro}  

\begin{figure} [ht]
   \begin{center}
   \begin{tabular}{lr} 
   \includegraphics[height=4cm]{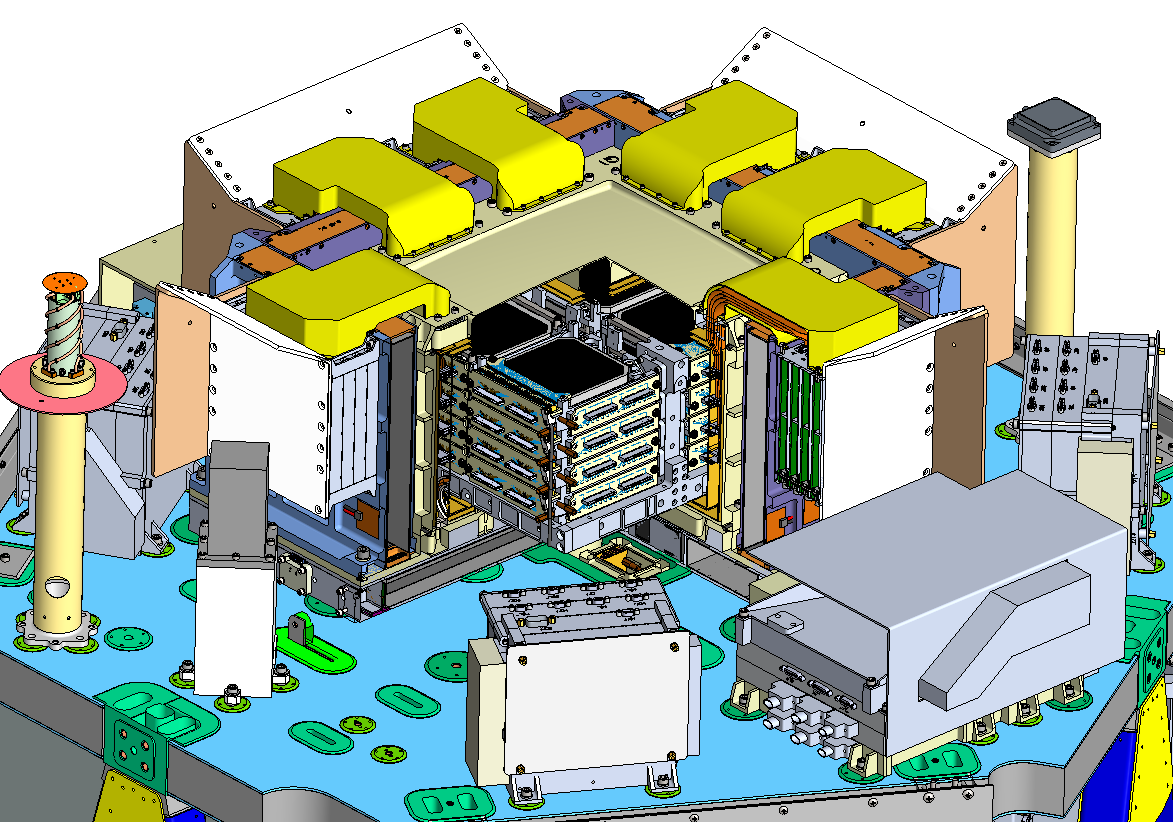}
     \includegraphics[height=4cm]{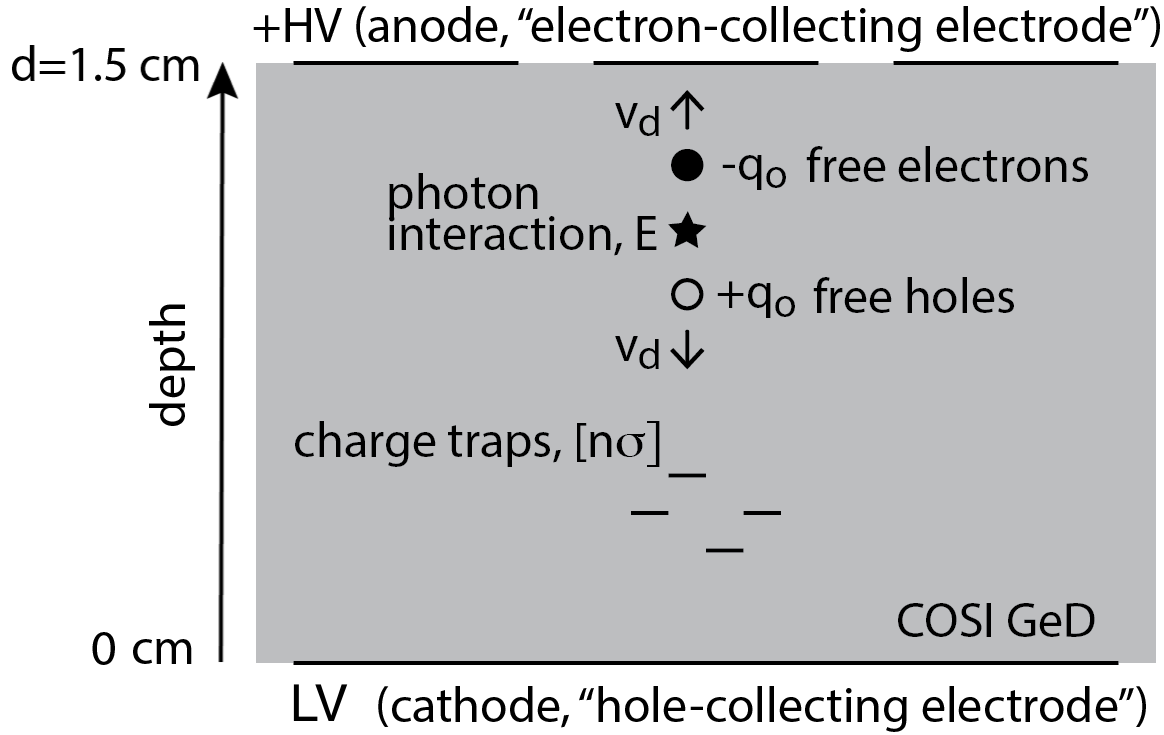}
   \end{tabular}
   \end{center}
   \caption[example] 
   { \label{fig:cosi} 
(left) Cutaway view of the COSI array\cite{tomsick2023comptonspectrometerimager}. (right) Diagram of the charge collection process in a COSI GeD induced by a photon interaction of energy E\cite{boggs2023numerical}. By measuring the time difference between collection of the free electrons and holes, we can measure the full 3D position of interactions within the cross-strip GeDs}
   \end{figure}

The Compton Spectrometer and Imager (COSI) is a wide-field gamma-ray (0.2-5 MeV) survey telescope designed to perform imaging, spectroscopy, and polarization of astrophysical gamma-ray sources \cite{tomsick2023comptonspectrometerimager}. It employs a novel Compton telescope design utilizing a compact array of 16 cross-strip germanium detectors (GeDs) (Fig. \ref{fig:cosi}) to resolve individual gamma-ray interactions with high spectral and spatial resolution. The COSI array is housed in a common vacuum cryostat cooled by a mechanical cryocooler. An active anticoincidence shield encloses the cryostat on the sides and bottom. The field-of-view of the instrument covers 25\% of the full sky at a given moment.

Coaxial high-purity germanium detectors have flown as high-resolution spectrometers on multiple space missions to study astrophysical sources \cite{vedrenne2003spi}, solar physics \cite{lin2002reuven}, and planetary compositions \cite{boynton2004mars, goldsten2007messenger, hasebe2009high}. All these previous space missions experienced degradation in the spectral performance of the detectors over time due to exposure to the space radiation environment. Annealing the detectors can repair radiation damage; however, the process of annealing risks damaging or even losing the operation of one or more of the GeDs from the process. Each of these previous missions utilized annealing to help maintain the spectral resolution of the GeDs with mixed success. To avoid the risk of losing the detectors, annealing for the planetary instruments was performed infrequently and at lower temperatures (85\textdegree \,C) than typical annealing (100\textdegree \,C), resulting in only partial repair of the radiation damage. RHESSI followed a similar approach of infrequent annealing, which allowed considerable degradation in the gamma-ray spectroscopy performance of the mission over its lifetime \cite{inglis2024spie}. INTEGRAL/SPI maintained a more aggressive approach of annealing the 19-GeD array every 6 months, which maintained the spectral resolution of the operational GeDs \cite{lonjou2005characterization} over the full mission lifetime, but potentially played a role in the failure of two detectors. 
 
Prior to our work, radiation damage effects in GeDs have been extensively studied for fast neutron induced damage due to predominantly neutron exposure in nuclear accelerator experiments \cite{kraner1968effects}. Proton induced damage, which dominates in the space environment, had been studied to a much lesser extent \cite{pehl1978high, koenen1995radiation}. However, both types of exposure produce disordered regions in the crystal structure that act as charge traps, leading to incomplete charge collection and resulting in degraded spectral performance. Radiation damage predominantly results in increased hole trapping in GeDs, with little effect on the electron charge collection \cite{kraner1968effects}. One way to limit the effects of radiation damage, which is not an option for COSI, is to implement an electrode geometry that minimizes the dependence on hole collection for the charge collection signal. For example, all the coaxial GeDs that have flown in space utilized ``reverse-electrode'' (n-type) geometries which have less sensitivity to hole trapping than ``conventional'' (p-type) coaxial GeDs \cite{pehl1979radiation}. 

In our program, we developed a physics-based numerical model of charge trapping effects in germanium (Sec. \ref{sec:model}), quantified the physical trapping parameters induced by proton radiation damage (Sec. \ref{sec:char}), developed and demonstrated an algorithm for correcting the effects of radiation damage on spectral performance of the COSI detectors (Sec. \ref{sec:corr}), and demonstrated the effectiveness of annealing in repairing damaged COSI detectors (Sec. \ref{sec:repa}). Here we will summarize these results, and discuss them in context of the impacts on COSI performance.

\section{MODELING CHARGE TRAPPING IN GERMANIUM}
\label{sec:model} 

\begin{figure} [ht]
   \begin{center}
   \begin{tabular}{c} 
   \includegraphics[height=5.8cm]{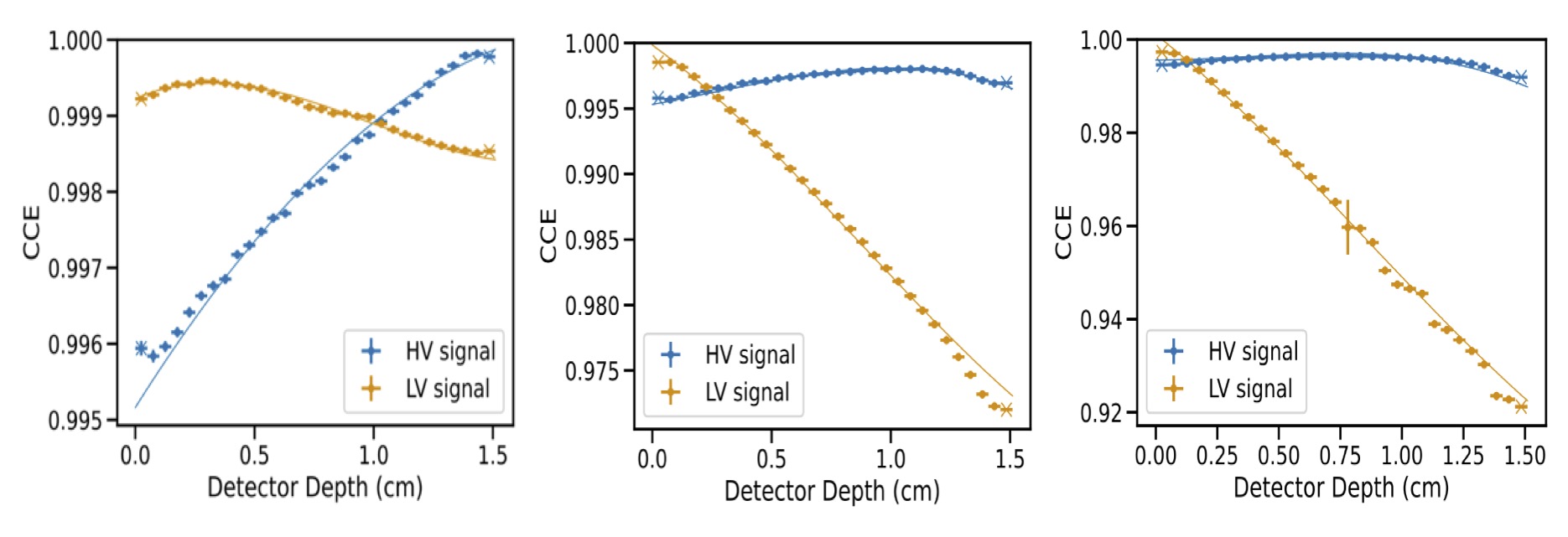}
   \end{tabular}
   \end{center}
   \caption[example] 
   { \label{fig:CCE} 
Charge Collection Efficiency (CCE) as a function of interaction depth measured with the COSI detector for electron-collecting strips (HV signal) and hole-collecting strips (LV signal) \cite{pike2025characterizing}. (left) Pre-damage curves, demonstrating the dominance of electron-trapping on the HV signal. (middle) After damage with a proton fluence of $2.0\times10^8 p^+/cm^2$, demonstrating the dominance of hole trapping on the LV signal after irradiation. (right) After extensive damage with a total proton fluence of  $5.0\times10^8 p^+/cm^2$. Shown for comparison (solid curves) are the corresponding best fit curves from our numerical charge trapping model \cite{boggs2023numerical}, which does an excellent job reproducing the measured curves. These curves were measured utilizing a Cs-137 (661.7 keV) source.}
   \end{figure}

Charge trapping is often characterized by either a mean free drift length, $\lambda$, which characterizes the distance a charge carrier can drift in the detector before being trapped; or alternately by mean trapping time, $\tau$, which characterizes the time that a charge carrier can drift in the detector before being trapped by a defect. Both of these models are approximations, valid in the limits that drift velocity $\gg$ thermal velocity ($\lambda$), or thermal velocity $\gg$ drift velocity ($\tau$).  We demonstrated that neither limit is valid for germanium, where drift velocities are comparable to thermal velocities, and so we introduced a model of trapping based on the fundamental trapping parameter, $[n\sigma]$ \cite{boggs2023numerical}, where $n$ is the density of traps and $\sigma$ is the associated cross section (Fig. \ref{fig:cosi}). The effects of trapping that we measure are due to the product $[n\sigma]$. This model is consistent with both $\lambda$ and $\tau$ in their respective velocity limits, and hence represents a universally general model for trapping in solid state detectors. We implemented and demonstrated this model in our own charge transport simulations and worked with the developers of the open-source Solid State Detectors package \cite{abt2021simulation} to implement our model of trapping in a recent release of the package. These numerical charge transport simulations were utilized to model the Charge Collection Efficiency (CCE) as a function of interaction depth in the strip detectors (Fig. \ref{fig:CCE}) for varying values of the electron and hole trapping parameters \cite{boggs2023numerical}. 

Trapping can vary significantly for holes ($[n\sigma]_h$) and electrons ($[n\sigma]_e$). Electron trapping is often assumed to be negligible, though we demonstrated this is not true for germanium detectors (Fig. \ref{fig:CCE}), and in fact intrinsic electron trapping dominates over hole trapping for undamaged detectors \cite{pike2023characterizing}. Radiation damage predominantly induces hole trapping, leading increased hole trapping for damaged detectors. For the COSI cross-strip detectors, the small pixel effect \cite{barrett1995charge} largely shields the electron collection signal from the effects of hole trapping. Thus, we anticipated, and subsequently demonstrated, that radiation damage will have a small impact on the spectral resolution of the AC/cathode/electron-collecting strips; however, the opposite is true for the hole-collecting strips, where the impact of the hole trapping is accentuated \cite{pike2025characterizing, HAIGHT2025170538}. 

\section{CHARACTERIZATION OF PROTON RADIATION DAMAGE IN GERMANIUM}
\label{sec:char} 

COSI will be launched into an equatorial Low Earth Orbit (LEO) (530 km altitude, 0\textdegree \, inclination), which is one of the most benign space radiation environments. The trapped proton fluence will be dominated by the passage of COSI through the edge of the South Atlantic Anomaly (SAA) for a few minutes each 90-minute orbit. The primary space radiation environment model used to estimate this trapped proton fluence is AP9, developed based on 45 satellite-based data sets \cite{ginet2013ae9}, which is available through ESA’s Space Environment Information System (SPENVIS) \cite{spenvis}. Utilizing AP9, we currently estimate an upper-limit on the 2-year exposure for COSI of $F_p \sim 1.1\times10^8 p^+/cm^2$ (95\% CL), and a best-estimate based on average fluences of $F_p \sim 0.6\times10^8 p^+/cm^2$ (CBE). We utilize these numbers for our estimated exposure times below. However, we note that comparison of AP9 with actual BeppoSAX particle monitor observations in equatorial LEO suggest that AP9 potentially overestimates the trapped proton fluences by an order of magnitude \cite{vripa2020comparison}. Regardless, we utilize the conservative AP9 as our upper limit on the trapped proton fluences.

We performed two proton beam irradiations of a spare COSI detector at the Loma Linda University (LLU) proton synchrotron. This GeD is nearly identical to the COSI flight detectors except for having 2-mm pitch electrode strips instead of 1.162-mm pitch like the flight detectors, which ultimately makes this spare GeD slightly more susceptible to the trapping effects on spectroscopy, but should have no effect on the radiation damage induced traps themselves. The LLU beamline can uniformly irradiate a 20 cm $\times$ 20 cm area with either a monoenergetic proton beam or a programmable proton spectrum (70-250 MeV). For our studies we utilized a monoenergetic beam of 150 MeV protons. The radiation damage due to these energetic protons should primarily depend on the total fluence, and not the detailed proton spectrum, due to protons above these energies being minimum ionizing particles (MIPs). We initially irradiated the GeD with a proton fluence of $F_p = 2.0\times10^8 p^+/cm^2$, corresponding to $\sim$ 4 years exposure (95\% CL) for COSI. After characterizing the performance after this initial exposure, we performed a second irradiation for a total proton fluence of $F_p = 5.0\times10^8 p^+/cm^2$, corresponding to ~10 years exposure (95\% CL) for COSI. These extreme exposures allowed us to characterize the relation between proton fluence and induced trapping, demonstrate our correction technique, as well as perform preliminary characterizations of the effectiveness of annealing on correcting this damage. 

While we anticipated that the radiation damage due to protons will scale proportionally to the total proton fluence, the scaling of $[n\sigma]_h$ on proton fluence $F_p$ has never previously been characterized in the literature. We are focused on the fluence of protons at energies $>$20 MeV as lower-energy protons will largely be stopped by passive materials surrounding the COSI detectors. Through our current work, we have been able to quantify the relation between energetic proton fluence and induced hole trapping product \cite{pike2025characterizing}:
\begin{equation}
\label{eq:fov}
[n\sigma]_h = (5.4\pm0.4)\times 10^{-11} F_p \;cm^{-1} \, .
\end{equation}
This scaling factor is an order of magnitude larger (i.e., more trapping) than an estimate for the dependence on neutron fluence $F_n$ \cite{raudorf1987effect}, which is consistent with empirical observations that radiation damage effects are detectable for much lower proton fluences than neutron fluences \cite{pehl1978high}. This is also consistent with the statement above that proton induced damage dominates in the space radiation environment.
	
In this work we also demonstrated the negligible effects of irradiation on electron trapping. The scaling factor between proton fluence and resulting hole trapping parameters was an unknown factor, crucial for modeling the impacts of on-orbit radiation exposure on the COSI detectors. Now that we have this relation between proton fluence and induced trapping, we can better investigate the impacts of the COSI exposure to the LEO space radiation environment. 

  \begin{figure} [ht]
   \begin{center}
   \begin{tabular}{c} 
   \includegraphics[height=8cm]{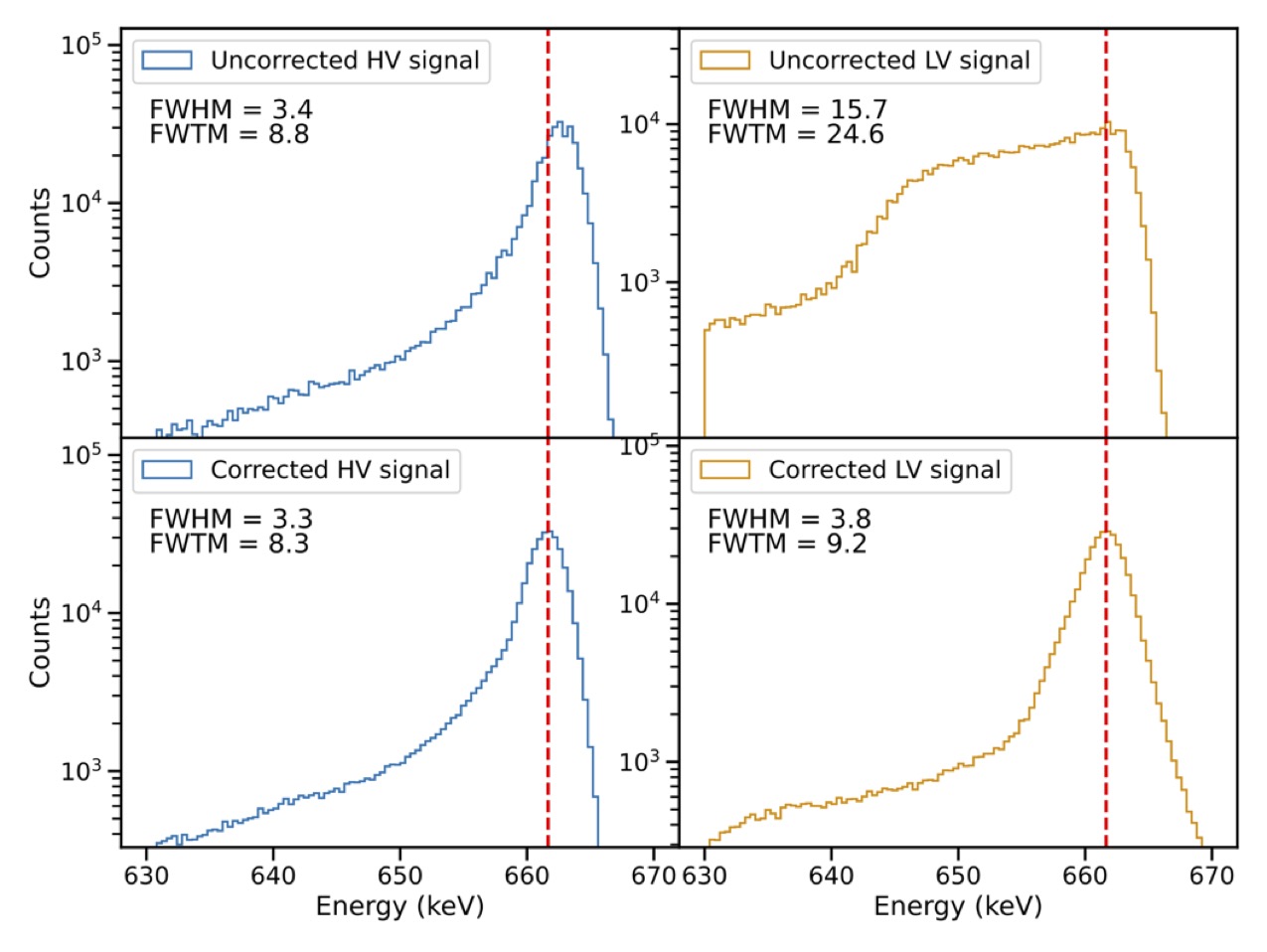}
   \end{tabular}
   \end{center}
   \caption[example] 
   { \label{fig:corr} 
Cs-137 (661.7 keV) spectra for the COSI detector after undergoing radiation damage with an energetic proton fluence of  $2.0\times10^8 p^+/cm^2$, corresponding to $\sim$4 years radiation exposure (worst case) in the COSI LEO radiation environment. (upper left) Electron-collection signal before correction, and (lower left) after correction. (upper right) Hole-collection signal before correction, and (lower right) after correction \cite{pike2025characterizing}. Our correction technique allows us to restore the spectral resolution of the hole-collecting signal to near pre-damage levels for the COSI detectors, even for this extreme level of radiation exposure. (The red dashed lines show the corrected peak positions for comparison between the spectra.)}
   \end{figure}

\section{CORRECTION OF TRAPPING EFFECTS ON SPECTRAL PERFORMANCE}
\label{sec:corr} 

COSI benefits from measuring the full 3D position of interactions within the cross-strip GeDs. The Charge Collection Efficiency (CCE) for a detector with intrinsic electron trapping and induced hole trapping will depend strongly on the depth of the interaction within the detector (Figs. \ref{fig:cosi}, \ref{fig:CCE}). We have demonstrated, with the COSI detectors, that by directly measuring this depth of interaction, we can apply a second-order depth-dependent correction to the energy calibration for each strip pair allowing us to correct for the effects of hole trapping and near-fully restoring the spectral resolution of the COSI detectors.

Before performing proton beam irradiations, we characterized the electron and hole trapping in several COSI detectors before undergoing radiation damage, demonstrating the dominance of intrinsic electron trapping in undamaged germanium \cite{pike2023characterizing}. We also demonstrated an empirical technique for correcting the effects of trapping using the depth of interaction, allowing us to improve the spectral resolution on the electron-collecting strip by ~30\%. 

We developed a physical model for charge trapping in germanium detectors 
(Sec. \ref{sec:model}), we integrated this model into our custom numerical charge transport simulations to model the CCE as a function of interaction depth in the strip detectors for varying values of the electron and hole trapping parameters. We demonstrated that we could calculate template CCE curves for a single value of the trapping parameters and then linearly scale these templates to accommodate any set of trapping parameters\cite{boggs2023numerical}. These templates provide a convenient method for fitting CCE curves to measured data in order to characterize the electron and hole trapping parameters and subsequently provide parameters for correcting the effects of trapping. These simulations were benchmarked against the measured CCE curves for an undamaged COSI detector, and the scaled CCE templates were utilized to successfully correct for the effects of intrinsic electron trapping.

After the GeD underwent extensive proton radiation damage, we demonstrated the degraded spectral performance as well as our ability to correct for this damage to restore the spectral resolution even for severe radiation damage \cite{pike2025characterizing} (Fig. \ref{fig:corr}).
We anticipate full recovery of the resolution for COSI detectors with moderate radiation damage.

\section{REPAIR OF RADIATION DAMAGE}
\label{sec:repa} 

\begin{figure} [ht]
   \begin{center}
   \begin{tabular}{c} 
   \includegraphics[height=6cm]{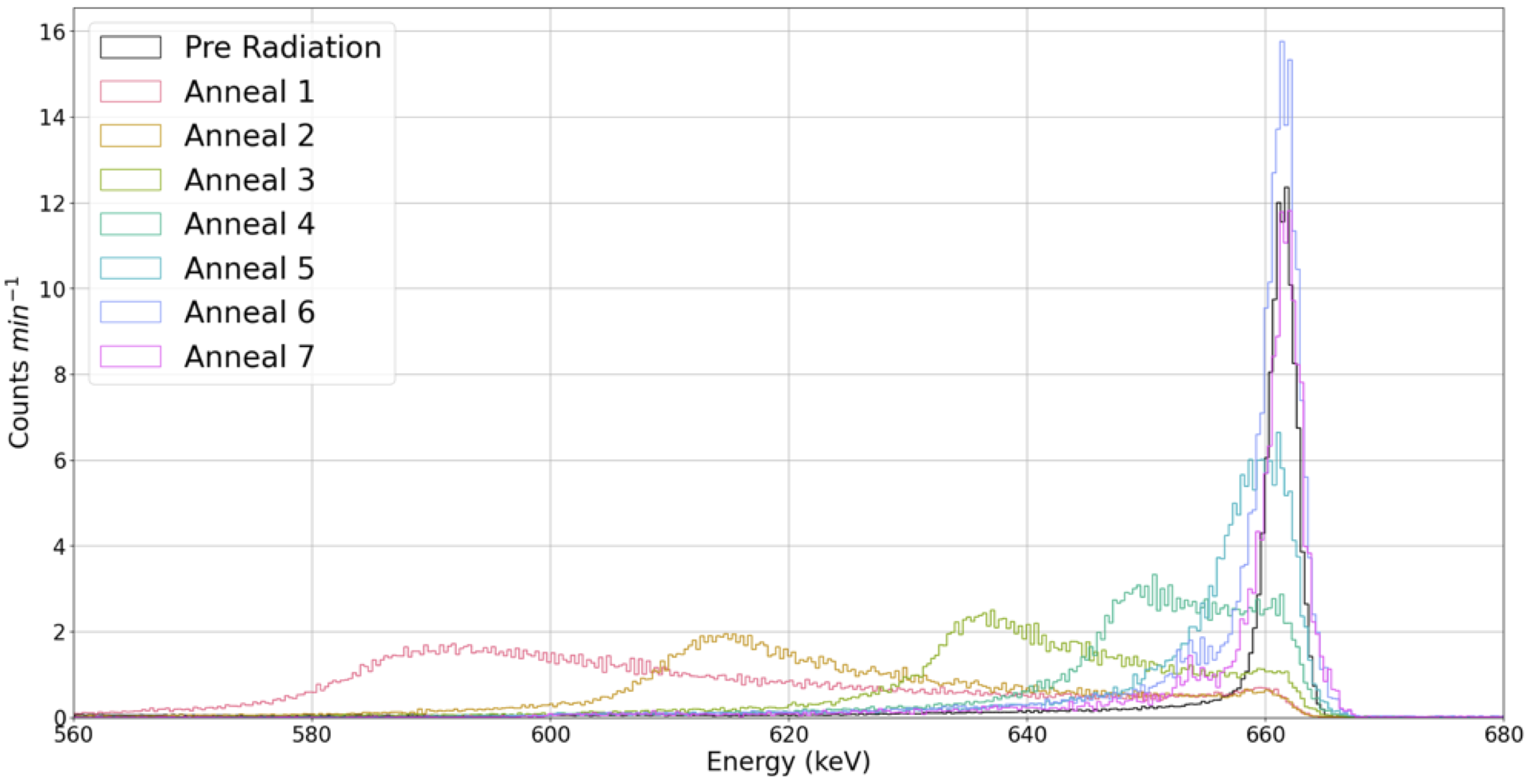}
   \end{tabular}
   \end{center}
   \caption[example] 
   { \label{fig:repa} 
The COSI LV (hole-collecting) signal Cs-137 (661.7 keV) spectra during different stages of high-temperature annealing, compared with the pre-radiation damage spectrum. Annealing was performed after the COSI detector was exposed to an energetic proton fluence of $5.0\times10^8 p^+/cm^2$, corresponding to $\sim$10 years radiation exposure (worst case) in the COSI LEO radiation environment. Anneals 1 \& 2 were performed at 80\textdegree C for a total of 72 hours and Anneals 3-7 were performed at 100\textdegree C for a total of 552 hours. The detector was cooled to 80K between anneals to test the spectral performance. The detector survived 7 annealing cycles, and the spectral resolution was ultimately restored to within 37\% of the pre-radiation resolution\cite{HAIGHT2025170538}.}
   \end{figure}

This correction technique will be less effective at higher levels of radiation damage. Hence, it is likely that COSI will need to perform high-temperature annealing to repair the radiation damage for an extended mission. As part of our program, we tested high-temperature annealing for survivability and effectiveness with the radiation damaged detector. We successfully demonstrated detector survivability for 7 high-temperature anneals, and were able to restore the spectral resolution to obtain a final FWHM of $4.08 \, keV$, within 37\% of the pre-radiation resolution ($2.98 \, keV$) even after extreme levels of radiation damage \cite{HAIGHT2025170538}(Fig. \ref{fig:repa}). We also characterized an exponential time scale of (118 $\pm$ 12 hr) for repair of the radiation damage, which predicts repair of he COSI detectors after the primary 2-year mission within $\sim$2 weeks for an annealing temperature of 100\textdegree C.

\section{DISCUSSION}
\label{sec:disc} 

There are several significant impacts to the COSI science that would result from degraded hole-collection signals. First, the optimal spectral resolution for COSI is achieved when the interaction energies are measured precisely on both the anode and cathode strips, which allows averaging of the electron- and hole-collection signals. Having one of the spectral resolutions significantly degraded removes this potential improvement in overall spectral resolution. Second, the electron-collection signals are on the AC-coupled side of the detectors, which typically has slightly poorer spectral resolution due to the extra coupling capacitance than the DC-coupled side. Thus, the hole-collection signals in undamaged detectors have the better spectral resolution. Third, matching the measured interaction energies on the two faces of the detector is an important factor in reconstructing the individual photon interactions from the measured raw data (charge collected and timing on individual triggered strips).

The correction technique that we have developed for germanium cross strip detectors will enable COSI to maintain its spectral resolution from launch through the full duration of its 2-year base-line mission without the need for high-temperature annealing. We anticipate, however, that annealing will be required to maintain the resolution for an extended mission. 

Going forward, we plan to perform proton beam irradiations on another spare GeD at more moderate proton irradiations, corresponding to 1- and 2-year exposures for COSI. These more moderate exposures will enable us to optimize our correction techniques and high-temperature annealing procedures for the COSI flight instrument.  
We are also studying the anticipated background spectral lines induced by SAA passages to develop methods of characterizing the radiation trapping parameters on orbit. These combined efforts will enable us to optimize the damage correction technique for COSI, track the radiation damage effects during the mission, as well as develop and verify annealing processes for the COSI detectors.

\acknowledgments 
 
This work was supported by the NASA Astrophysics Research and Analysis (APRA) program, grant 80NSSC22K1881. The Compton Spectrometer and Imager is a NASA Explorer project led by the University of California, Berkeley with funding from NASA, United States under contract 80GSFC21C0059. This work was supported by the Laboratory Directed Research and Development Program of Lawrence Berkeley National Laboratory under U.S. Department of Energy Contract No. DE-AC02-05CH11231. We would like to acknowledge the staff at the James M. Slater, MD, Proton Treatment and Research Center, Loma Linda University Medical Center beamline for their assistance in this study. 

\bibliography{refs} 
\bibliographystyle{spiebib} 

\end{document}